\documentclass[11pt, a4paper]{article} 

\usepackage[includeheadfoot, width=442pt, height=720pt]{geometry}
\usepackage{amsmath}
\usepackage{amsfonts}
\usepackage{amssymb}
\usepackage{bbm,bm}
\usepackage{mathtools}
\usepackage{slashed}
\usepackage{mathalfa}
\usepackage{graphicx,caption,subfigure}
\usepackage{tikz} 
\usepackage{tikz-cd}
\usetikzlibrary{decorations.pathreplacing,calc,tikzmark}
\usepackage{booktabs}
\usepackage{adjustbox}
\usepackage[utf8]{inputenc}
 \usepackage[splitrule]{footmisc}

\usepackage{placeins}
\usepackage{empheq}
\usepackage{paralist}
\usepackage{cite}
\usepackage[normalem]{ulem}
\usepackage{soul}

\usepackage[colorlinks=false,urlbordercolor=red,linktoc = page]{hyperref}
\usepackage{xcolor}
\usepackage{tensor}

\allowdisplaybreaks[2]

\DeclareUnicodeCharacter{2212}{-}

\begin{document}

\phantom{.}

\vspace{1.5cm}
\begin{center}
 {\Large \bf Scale separation, rolling solutions  and entropy bounds}
\vspace{0.6cm}

\end{center}

\begin{center}
{\large
David Andriot$^a$, Niccol\`o Cribiori$^b$, Thomas Van Riet$^{b}$
}
\end{center}

\vspace{0.1cm}
\begin{center} 
\emph{
$^a$Laboratoire d’Annecy-le-Vieux de Physique Th\'eorique (LAPTh),\\
CNRS, Universit\'e Savoie Mont Blanc (USMB), UMR 5108,\\
9 Chemin de Bellevue, 74940 Annecy, France \\
 \vspace{0.3cm}
$^b$ KU Leuven, Institute for Theoretical Physics,\\ Celestijnenlaan 200D, B-3001 Leuven, Belgium \\[.1cm] 
    } 
\end{center} 

\vspace{2.0cm}

\begin{abstract}

We revisit scale separation for compactifications of ten- and eleven-dimensional supergravity. For cosmological solutions rolling down flux-generated potentials, we observe that scale separation is achieved as time flows, and is fairly generic. This is realized without the need of orientifolds nor corrections to the classical supergravity approximation. We then confront scale separation with the Covariant Entropy Bound (CEB) and the CKN bound. We show that a naive application of these bounds to vacua hints at the existence of at least two extra dimensions. For rolling solutions, we observe that the CEB is not always respected, but since these examples lack a cosmic horizon, the application of entropy bounds remains delicate.

\end{abstract}

\thispagestyle{empty}
\clearpage

\setcounter{tocdepth}{2}

\tableofcontents

\section{Introduction}

The Electroweak and the Cosmological hierarchy problems are important clues to understand the physics beyond the Standard Model if one accepts that they should have an (non-anthropic) explanation. 
In a modern Swampland perspective \cite{Vafa:2005ui, Palti:2019pca,Agmon:2022thq}, both hierarchy problems indicate that the universe is in some loose sense in an ``asymptotic regime of moduli space''. 
This entails the exciting possibility that Swampland principles are of direct relevance to our observed universe, as they become genuinely constraining in those regimes. In essence, the smallness of the Electroweak scale and of the dark energy scale predicts dark light sectors. Note that this happens differently from how usual naturalness arguments predict light sectors, say, as in the Large Extra Dimension Scenarios. For the Cosmological hierarchy problem the appearance of a light tower is anchored to what is known as the ``AdS Distance Conjecture'' (ADC) \cite{Lust:2019zwm}, and to its putative extension to positive energy backgrounds. Similarly, light towers have been proposed whenever Yukawa couplings become small such that light fermions appear \cite{Palti:2020tsy, Cribiori:2020use, Gonzalo:2021fma, Cribiori:2021gbf, Castellano:2021yye, DallAgata:2021nnr, Gonzalo:2021zsp, Cribiori:2023gcy, Casas:2024ttx}; such towers could thus appear in the limit where the BEH field vanishes. Starting from the ADC, applied to the observed universe, \cite{Montero:2022prj} concluded that there must be one extra dimension of mesoscopic scale, which is known as the \emph{Dark Dimension Scenario}.

In the context of ultraviolet completions of the Standard Model with extra dimensions, like in string theory, the Cosmological hierarchy problem is essentially identical to what is referred to as \emph{scale separation} of a vacuum  solution, such as $\mathcal{M}_4\times\mathcal{M}_6$.\footnote{With ``vacuum'' we refer to a maximally symmetric solution of the equations of motion. In contrast to a rolling solution, it can correspond to an extremum of the scalar potential. \label{footnote:vacuum}} Scale separation means that the curvature radius (the cosmological constant or the Hubble scale), $L_H$, of the four-dimensional spacetime $\mathcal{M}_4$ is significantly larger than the Kaluza-Klein length scale, $L_{KK}$, associated to $\mathcal{M}_6$, $L_{KK}L^{-1}_H\ll 1$. This condition is not easy to meet in vacua with stabilised moduli, as for instance reviewed in \cite{Coudarchet:2023mfs}. On the other hand, it is needed to have a four-dimensional (or low-dimensional) effective description of the fluctuations around the vacuum. Cosmological hierarchy usually refers to $L_{H}$ being large in units that set the scale of new physics, so certainly $L_{KK}$ serves that purpose. Interestingly, this means that a Cosmological hierarchy is always occurring in vacua that appear four-dimensional to an observer in a universe described by critical string theory \cite{Gautason:2015tig}. In this sense, the two concepts, scale separation and Cosmological hierarchy, are identical.

The no-go theorem of \cite{Gautason:2015tig} provides a pragmatic reason as to why scale separation in string/M-theory is challenging. In the realm of two-derivative effective actions with solely classical ingredients, and under the assumption that the typical size of the extra dimensions cannot be decoupled from their integrated curvature, the no-go theorem states that scale-separated vacua require negative tension singularities, $i.e.$ orientifolds.\footnote{To the best of our knowledge, the only construction evading this no-go theorem without orientifolds and with only classical ingredients has been proposed in \cite{Cribiori:2021djm, VanHemelryck:2024bas}. These scale-separated anti-de Sitter vacua of eleven-dimensional supergravity \`a la Freund-Rubin evade \cite{Gautason:2015tig} because the Kaluza-Klein scale is decoupled from the compact space integrated curvature. In ten dimensions, such decoupling also occurs on nilmanifolds \cite{Andriot:2018tmb}. \label{footnote1}}  The hypothesis of this no-go theorem are identical (up to the assumption mentioned) to the Maldacena-Nunez (MN) no-go theorem \cite{Maldacena:2000mw} that forbids Minkowski and de Sitter vacua. Similar to the work-around of the MN no-go, orientifolds seem key for achieving scale separation; see nevertheless footnote \ref{footnote1}.

Importantly, it is known that the MN no-go theorem can also be evaded by allowing rolling scenarios, by which we mean a time-dependent solution with scalar fields rolling along a positive potential. Then,  cosmic acceleration, even if not in the form of de Sitter, is realizable once time-dependent extra dimensions are allowed \cite{Townsend:2003fx, Emparan:2003gg}. The first goal of this note is to point out that, similarly, rolling scenarios can be a way out of the no-go of \cite{Gautason:2015tig} against scale separation. The fact that rolling solutions can be scale-separated has been mentioned in passing in \cite{Rudelius:2022gbz, Shiu:2023fhb, Andriot:2023wvg}. We believe this to be an important point, especially in view of a Swampland perspective that suggests that small couplings or large hierarchies are generic in asymptotic regions where the potential becomes runaway \cite{Ooguri:2018wrx}. It could also be important in view of realizing a dynamical dark energy with such solutions from string theory, as recently suggested by observations \cite{DESI:2024mwx, DESI:2025zgx}. We will exemplify this point in section \ref{sec:vacuavsroll}, where we will show how a potential purely generated from internal curvature cannot help in finding scale separation \cite{Shiu:2023fhb}. Different is the case of a flux-supported scalar potential, for which we provide top-down constructions giving scale separation with solely classical ingredients. The way scale separation occurs in these constructions seems to be fairly generic.

The second goal of this note involves entropy bounds, presented in section \ref{sec:entropy}. These can also be used to constrain extra dimensions, as already done in \cite{Castellano:2021mmx,  Castellano:2023qhp, Calderon-Infante:2023ler}. We will apply the covariant entropy bound \cite{Bousso:1999xy} and the stronger CKN bound \cite{Cohen:1998zx} to both rolling and vacuum solutions. We follow the logic of \cite{Castellano:2021mmx}, but remain open minded about the possibility that a more careful analysis is needed. In this spirit, we point out that the covariant entropy bound, in our naive implementation, is violated by the Dark Dimension scenario \cite{Montero:2022prj} when taken as a vacuum, but it can be compatible with the existence of two (dark) extra dimensions \cite{Casas:2024ttx, Anchordoqui:2025nmb} if one does not insist on the relation $L_{KK} \sim \Lambda^{-1/4}$, with $\Lambda$ the cosmological constant. The rolling cosmological solutions we consider do not have an apparent horizon, and it is unclear whether we can apply entropy bounds using the Hubble radius as an IR scale.\footnote{By now there is ample circumstantial evidence for not having cosmic horizons without going beyond the classical two-derivative action without orientifolds. More profoundly, it could even be that cosmic acceleration from quantum gravity can never be sufficient to allow for a cosmic event horizon \cite{Andersson:2006du, Andriot:2023wvg}.} If we nonetheless do so, we find that the solutions do not always respect the covariant entropy bound, while CKN is even more difficult to satisfy. We discuss and interpret these results in section \ref{sec:disc}.

\section{Scale separation: vacua versus rolling solutions}
\label{sec:vacuavsroll}

Our starting point is the no-go theorem of  \cite{Gautason:2015tig}. This is basically an extension of the celebrated no-go \cite{Gibbons:1984kp,deWit:1986mwo,Maldacena:2000mw} forbidding Minkowski and de Sitter vacua (see footnote \ref{footnote:vacuum}) in ten-dimensional supergravity, if only classical ingredients are present and negative tension objects are absent. The main observation of  \cite{Gautason:2015tig} is that, if the Kaluza-Klein scale cannot be decoupled from the internal curvature, the very same reasoning of \cite{Gibbons:1984kp,deWit:1986mwo,Maldacena:2000mw} forbids also scale-separated anti-de Sitter vacua. In this section, we investigate how such no-go theorems are evaded by looking instead for \emph{rolling solutions}. Crucially, our examples will be purely classical and without negative tension objects.

\subsection{The no-go theorem for vacua}

Let us first give a simplified, alternative derivation of the no-go theorem of \cite{Gautason:2015tig} using a four-dimensional perspective. We consider an effective theory coming from a compactification of type II string theory taking the form
\begin{equation}
\label{S4d}
{\cal S} = \int d^4 x \sqrt{|g_4|} \left( 
\frac{M_P^2}{2} \mathcal{R} -\frac{1}{2} \sum_i(\partial\varphi^i)^2 -V \right) \,,
\end{equation}
in four-dimensional Einstein frame. Below we  set $M_P=1$.  
We include $p$-form RR fluxes and 3-form NSNS flux threading the extra dimensions, curvature contribution(s) from the internal space, and (smeared) $D(3+k)$-branes wrapping internal $k$-dimensional cycles.\footnote{We do not consider $NS5$-branes given the difficulty in solving the associated tadpole. Similarly, we do not consider $D$-branes with co-dimension less than three, since their backreaction typically leads us away from a controlled regime; naively including them does not change the no-go theorem, up to an order one constant.} 
We also assume the absence of non-trivial warp factors and dilaton gradients; this is often the case with smeared sources.
Among the scalar fields we have the volume $\rho^3$ (in string units) of the compact six-dimensional space and the dilaton $\phi$.
In terms of the universal scalars $\rho$ and $\tau=\rho^{3/2} e^{-\phi}$, the four-dimensional potential reads \cite{VanRiet:2023pnx}
\begin{equation}
\begin{aligned}
\label{eq:V4dnogo}
V&= \sum_{p=0}^6 U_p\ \rho^{3-p} \tau^{-4} + U_{H_3}\  \rho^{-3}\tau^{-2} + \sum_{k=0}^3 U_k \ \rho^{\frac{k-3}{2}}\tau^{-3} + U_{R_6} \ \rho^{-1}\tau^{-2}\,,\\
&\equiv  \sum_p V_p + V_{H_3} + \sum_k V_k + V_{R_6}\,,
\end{aligned}
\end{equation}
from which one obtains the following expressions
\begin{equation}
\begin{aligned}
V &= - \sum_p V_p - \frac{1}{2}\sum_k V_k - \frac{1}{2} \tau \partial_{\tau} V \ , \\
V_{R_6} &= -\sum_p \frac{9-p}{2} V_p - \sum_k \frac{k+6}{4} V_k - \frac{3}{4} \tau \partial_{\tau} V + \frac{1}{2} \rho \partial_{\rho} V \ . \label{VpartialV} 
\end{aligned}
\end{equation}
The key assumption of \cite{Gautason:2015tig} is that the Kaluza-Klein scale is not decoupled from the internal curvature. In the above language this means, on a vacuum, that $L_{KK}^{-2} \simeq |V_{R_6}|$. Similarly, on a vacuum we have ${\cal R} = 4 V=4 \Lambda$ in four-dimensional Planck units, so we define the four-dimensional scale $L_H^{-2} \simeq |V|$, up to order-one factors. On a vacuum, we necessarily have $\tau \frac{\partial V}{\partial \tau}=0=\rho \frac{\partial V}{\partial \rho}$. From \eqref{VpartialV}, we then arrive at
\begin{equation}
\begin{aligned}
\frac{L_{H}^2}{L_{KK}^2} &\simeq \left| \frac{V_{R_6}}{V}\right| = \left|\frac{ \sum_p\left(\frac{9-p}{2}\right) V_p +\frac12 \sum_k \left(\frac{k+6}{2}\right)V_k}{\sum_p V_p + \frac{1}{2}\sum_k V_k}\right|\leq \frac 92\,,
\end{aligned}
\end{equation}
where we used that all terms $V_p, V_k$ are positive. Hence, within the assumptions made, we find an obstacle to scale separation. 

On a vacuum, the way out is to allow for negative contributions. On physical grounds, $V_p$ must be positive since they are squares of fluxes. $V_k$ is the contribution of $(3+k)$-branes, which we assumed to be $D$-branes and in such a case $V_k>0$. However, orientifolds would contribute precisely with the same dependence on $\rho$ and $\tau$, except for the fact that $U_k<0$ giving $V_k<0$. Hence, within previous assumptions, orientifolds are necessary (but not sufficient)  to achieve scale separation. Another way to circumvent this no-go theorem is to relax the assumption that $L_{KK}^{-2} \simeq |V_{R_6}|$, as in \cite{Cribiori:2021djm,VanHemelryck:2024bas}, or by moving away from vacua, say to rolling solutions. Indeed, one can see from \eqref{VpartialV} that having fields on a non-vanishing potential slope would prevent one from reaching the above conclusion. In the following, we discuss this latter possibility.

\subsection{Rolling away from the no-go theorem}

Let us again work in four dimensions with the action \eqref{S4d} although the analysis can be generalized to compactifications to arbitrary dimensions. On a de Sitter vacuum, there is only one four-dimensional length scale $\mathcal{R} =4 V =4 \Lambda = 12 H^2$. In addition, the Hubble length $H^{-1}$ is related to the size of the cosmological event horizon. We then typically use as infrared cutoff the Hubble scale $L_H$ defined as $L^{-2}_H \simeq |\mathcal{R}| $. To go beyond vacuum solutions, we consider in the following a cosmological FLRW spacetime with spatial curvature $k$. 

To discuss scale separation, one has to understand what lengths should be compared. Regarding the four-dimensional one, in analogy to the vacuum case we consider again $\mathcal{R}$ in four-dimensional Einstein frame, where now
\begin{equation}
\mathcal{R} = 6 \left( H^2 + \frac{\ddot{a}}{a} + \frac{k}{a^2} \right) \ .
\end{equation}
Here, $a(t)$ is the scale factor and a dot represents a derivative with respect to cosmic time, while $H= \frac{\dot{a}}{a}$. With solutions of the form $a(t) \sim t^s$ and $k=0$, one gets
\begin{equation}
\mathcal{R} = 6 \left(2 - \frac{1}{s} \right) \, H^2 \ .
\end{equation}
Hence, we can pick again the Hubble scale as the relevant four-dimensional scale,
\begin{equation}
L_H\simeq H^{-1} \sim t\, .    
\end{equation}
Another option could have been to use instead $V^{-1/2}$ as a four-dimensional length scale, as proposed in \cite{Andriot:2022brg} (see also \cite{Rudelius:2022gbz}). For the solutions to be discussed below, one gets $V(\varphi(t))^{-1/2} \sim t \sim L_H(t) $, therefore giving the same results.

Next, we turn to the length scale of the internal space. We consider a $(4+n)$-dimensional Einstein frame metric of the form
\begin{equation}
\label{4+nmetric}
d \tilde{s}_{4+n}^2 = d \tilde{s}_4^2 + d \tilde{s}_n^2 = e^{2 A \varphi}\, d s_4^2 + e^{2 B \varphi}\, d s_n^2 \ ,
\end{equation}
where $d s_4^2$ will eventually be the four-dimensional Einstein frame metric under dimensional reduction. The dependence on $\varphi$ is introduced for later convenience, but is not relevant in the following reasoning. For $n=6$ one reads a Kaluza-Klein length $\sim e^{B \varphi}$ in ten-dimensional Planck units, to be converted to four-dimensional Planck units via the standard relation $M_{P,10} \sim M_{P}\,  e^{-3 B \varphi}$. It is instructive to see how this works in practice, not necessarily on a vacuum. Consider the $(4+n)$-dimensional gravity action coupled to a scalar field $\Phi$. Neglecting derivatives on $\varphi$ for simplicity, we get in Einstein frame
\begin{align}
&\int d^4 x \sqrt{|\tilde{g}_4|} \int d^n x \sqrt{|\tilde{g}_n|} \ \frac{1}{2} \left( \tilde{{\cal R}} + \tilde{{\cal R}}_n - (\partial \Phi)^2_{4+n} \right) \label{action4n}\\
& = \int d^4 x \sqrt{|g_4|} \int d^n x \sqrt{|g_n|}\ e^{(2A+nB) \varphi} \ \frac{1}{2}\left( \ {\cal R} + e^{2(A- B) \varphi}\ {\cal R}_n -  (\partial \Phi)^2_4 + e^{2(A-B) \varphi}\ \Phi \Delta_n \Phi \right)\ ,\nonumber
\end{align}
where we extracted all $A,B$ dependence. In order to reach the four-dimensional Einstein frame, we need to pick $2A+nB=0$. Using the standard Kaluza-Klein mass generation, $\Delta_n \Phi = -m^2 \Phi$, we end up with the four-dimensional action
\begin{equation}
\int d^4 x \sqrt{|g_4|}
\int d^n x \sqrt{|g_n|} \ \frac{1}{2}
\left(  {\cal R} + e^{-(2+n) B \varphi}\ {\cal R}_n - (\partial \Phi)^2_4 - e^{-(2+n) B \varphi}\ m^2 \Phi^2 \right)\ , \label{finalactiondimred}
\end{equation}
where the overall constant internal volume is usually absorbed into $M_P$. 
From the above, one can read-off the four-dimensional Klein-Gordon equation and hence the Kaluza-Klein mass. We thus see that in four-dimensional Einstein frame, both the Kaluza-Klein mass squared and the internal curvature scale as $e^{-(n+2) B \varphi}$, and not as the naive $e^{-2 B \varphi}$ that one would read from the $(4+n)$-dimensional Einstein frame metric \eqref{4+nmetric}. The internal length scale to consider is then
\begin{equation}
L_{KK} \sim e^{\frac{n+2}{2} B \varphi} \ , \label{LKK}
\end{equation}
in four-dimensional Planck units. This is to be compared with $L_H \sim H^{-1}$ arising from the four-dimensional Einstein frame curvature ${\cal R}$.

Having defined the proper scales to study scale separation on rolling solutions, we now revisit remarks made in \cite{Shiu:2023fhb, Andriot:2023wvg}. It was noted there that runaway solutions can achieve scale separation if along the time flow
\begin{equation}
\frac{L_{KK}(t)}{L_{H}(t)}\sim t^{P},\qquad \text{with} \qquad P<0    \,.
\end{equation}
We illustrate this idea in the following examples and discuss the consequences; crucially, we will see that its implementation has some subtleties. 

\subsubsection*{Single-field exponential}
Consider the four-dimensional theory of gravity \eqref{S4d} coupled to a single scalar field $\varphi$, with potential
\begin{equation}
\label{example1}
 V(\varphi) = V_0\ e^{-\lambda \varphi} \ , 
\end{equation}
where $V_0 > 0$ and we set $M_P=1$.\footnote{For a recent and complete overview of cosmological solutions with an exponential potential, see \cite{Andriot:2024jsh}.} The FLRW metric is denoted $d s_4^2$, with Ricci scalar ${\cal R}$. For $k=0$, this theory admits a late time cosmological attractor solution which depends on $\lambda$ \cite{Lucchin:1984yf},
\begin{align}
\label{l2<6}
&\lambda^2 \leq 6\,\, ,\quad a(t)\sim t^{2/\lambda^2},\,\quad \varphi(t) =\frac{2}{\lambda}\log t + c ,\\
&\lambda^2>6\,\, , \quad a(t)\sim t^{1/3},\,\quad  \varphi(t) =(sign{\lambda})\sqrt{\frac 23}\log t + c',\label{kination}
\end{align}
where the constants $c, c'$ are not important for our discussion. 

With negative spatial curvature ($k=-1$) in the external space, the late time attractor is \cite{vandenHoogen:1999qq}
\begin{equation}
\lambda^2 > 2\,\, , \quad a(t) \sim \, t \ ,\,\quad \varphi(t) = \frac{2}{\lambda} \log t + c, \label{solPkp}
\end{equation}
where $c$ is a constant not relevant for our purposes. We will now consider examples from compactifications of eleven- and ten-dimensional supergravity that lead to single exponential potentials with different values of $\lambda$. We then uplift the attractor solutions described above and verify whether scales can be separated. 

\subsubsection*{Compactifications of eleven-dimensional supergravity}

The first example we consider is a compactification of eleven-dimensional supergravity on a seven-dimensional compact hyperboloid without flux. Following \cite[(3.5)]{VanRiet:2023pnx}, the eleven-dimensional Einstein frame metric ansatz is
\begin{equation}
ds^2_{11} = e^{2A\varphi}ds^2_4 + e^{2B\varphi}ds^2_7\,,    \label{11dmetric}
\end{equation}
with  $2A=-7B$ and $A^2 = 7/18$, leading to \eqref{example1} with $2<\lambda^2 = \frac{18}{7}<6$. 
Therefore, both the $k=-1$ and $k=0$ attractors have $e^{-9B\varphi}=e^{-\lambda \varphi} \sim 1/{t^2}$ giving from \eqref{LKK} a Kaluza-Klein scale $L_{KK} \simeq e^{\frac{9}{2} B \varphi} \sim t$. Eventually, we find
\begin{equation}
\frac{L_{KK}}{L_H}\sim t^{0} \ ,
\end{equation}
meaning there is no scale separation occurring as time flows. Circumventing the logic of \cite{Gautason:2015tig} by going away from a vacuum solution is therefore not enough in this example.\footnote{Nevertheless, from an eleven-dimensional viewpoint it is a vacuum solution. The lift of the four-dimensional FLRW solution with $k=0$ to eleven dimensions corresponds to an eight-dimensional Milne metric times the Euclidean three-dimensional plane. As such, it has vanishing Riemann curvature. Yet, the Milne metric, $ds^2 =-dt^2 + t^2ds^2_7$, only corresponds to eight-dimensional Minkowski space if $ds^2_7$ is the \emph{non-compact} hyperboloid.}
This can readily be understood from the observation that the scalar potential comes from internal curvature, as noticed in \cite{Shiu:2023fhb}. Indeed, we showed that a (non-zero) internal curvature scales as the Kaluza-Klein mass square, so that $L_{KK} \sim |V_{{\cal R}_n}|^{-1/2}$. We also observed that for the solutions considered, $L_H \sim V(\varphi(t))^{-1/2}$. If the potential $V$ is generated by ${\cal R}_n$ only, $i.e.$~$V \sim V_{{\cal R}_n}$, both lengths have the same scaling.\footnote{
Indeed, if the scalar potential is generated solely by internal curvature, from \eqref{finalactiondimred} we read that $V(\varphi)^{-1/2} \sim L_{KK}$, and $V$ is as in \eqref{example1} with $\lambda=2(B-A)= (n+2) B = - 2\frac{n+2}{n} A$. To get a canonical kinetic term for $\varphi$, we set $2A^2 (n+2)=n$, see $e.g.$ \cite[(3.5)]{VanRiet:2023pnx}. We then find $\lambda^2 = 2+4/n$, implying $2 < \lambda^2 \leq 6$ for $n\geq 1$ and thus leading to the solution \eqref{l2<6} or \eqref{solPkp} as a late time attractor. As a consequence, $L_H \sim t  \sim  e^{\frac{\lambda}{2}\varphi} \sim V(\varphi)^{-1/2} \sim L_{KK}$.}

As a second example, let us consider eleven-dimensional supergravity on a Ricci flat $7$-torus with $7$-form flux (magnetic dual to an external $4$-flux). The Einstein frame metric is \eqref{11dmetric}. The scalar potential is now generated by the flux and is a single exponential as in \eqref{example1}, with $\lambda=21B$ and $\lambda^2 =14$. 
For $k=0$ the late time attractor solution is given by \eqref{kination}, and from \eqref{LKK} we find $L_{KK} \sim t^{\sqrt{3/7}}$, such that 
\begin{equation}
\frac{L_{KK}}{L_H} \sim t^{-1+\sqrt{3/7} }\rightarrow 0 \qquad \text{for} \qquad t \to \infty,    
\end{equation}
and so this solution is scale-separated as time flows. For $k=-1$, the solution corresponds to the attractor \eqref{solPkp}. We then get $e^{-\lambda\varphi} \sim \frac{1}{t^2}$ and, from \eqref{LKK}, $L_{KK} \sim t^{\frac{3}{7}}$, such that
\begin{equation}
\frac{L_{KK}}{L_H} \sim t^{-4/7 }\rightarrow 0 \qquad \text{for} \qquad t \to \infty,
\end{equation}
and so is again scale-separated. Crucially, this works thanks to a potential \emph{not generated by internal curvature}.

\subsubsection*{Compactifications of ten-dimensional type II supergravity}

To construct more examples with scale separation,\footnote{Another non-scale-separated example is obtained as a consistent truncation of type IIA supergravity compactified on a negatively-curved six-dimensional Einstein manifold \cite{Andersson:2006du,Marconnet:2022fmx, Andriot:2023wvg}. The curvature-generated potential has $\lambda=\sqrt{8/3}$. For $k=-1$ or $k=0$ this gives at late time $L_{KK} \simeq e^{\frac{2}{\sqrt 6}\varphi} \sim t$, hence no scale separation.} let us focus on the following part of the ten-dimensional type II supergravity action in Einstein frame
\begin{equation}
S \supset \int d^{10} x \, \sqrt{|g_{10}|}\left( \frac{1}{2} {\cal R}_{10} -\frac{1}{2}(\partial \Phi)^2 - \frac{1}{4}\, e^{a_m\Phi}\, \frac{{F_m}^{\!\!2}}{m!} \right) \,,   
\end{equation}
with $\sqrt2\, \Phi=\phi$ where $\phi$ is the dilaton, $a_m =\frac{5-m}{\sqrt2}$ and $F_m$ are RR $m$-form field strengths. In the following we will consider $m=0,1,\dots,6$.\footnote{While fundamentally one should stop at $m=4$ in type IIA, and $m=5$ with a different flux prefactor in type IIB, considering magnetic duals to fluxes with external components allows to consider as well the above flux term with $m=5,6$. This non-trivial rewriting was described for instance in \cite[App.A]{Andriot:2020lea}.} 
More general setups have been recently studied in \cite{Casas:2024oak}. A compactification on a six-dimensional Ricci-flat compact manifold with metric $ds^2_6$ is described in ten-dimensional Einstein frame by
\begin{equation}
    ds^2_{10}= e^{2A\varphi}ds^2_4 + e^{2B\varphi}ds^2_6\,,
\end{equation}
where $ A=-3B$, and the canonically normalized volume gives $A^2 =3/8$. The four-dimensional action in Einstein frame is then
\begin{equation}
S_4\supset   \int d^4 x\, \sqrt{|g_{4}|}\left(\frac{1}{2} {\cal R} - \frac{1}{2}(\partial \Phi)^2 - \frac{1}{2}(\partial \varphi)^2 - V \right)\,, 
\end{equation}
with the potential $V$ generated by $F_m$.

To get a single exponential potential as a consistent truncation we can distribute the flux in an isotropic way on the internal space. The resulting flux energy-momentum tensor is then proportional to the metric. For instance, if we consider a torus with Cartesian coordinates $\theta^1,\ldots, \theta^6$ then such flux choice could be of the type
\begin{align}
 & F_2 \sim N\left( d\theta^1\wedge d\theta^2 + d\theta^3\wedge d\theta^4  + d\theta^5\wedge d\theta^6 \right)\,,
\end{align}
with $N$ the quantised flux quantum. Many other options exist and they all have in common that one can truncate all deformations of the torus aside from the volume. While the compact, Ricci flat manifold need not be Calabi-Yau, in the latter case one can choose $F_2\sim J$ with $J$ the K\" ahler 2-form, as in the example of \cite{Marconnet:2022fmx, Andriot:2023wvg}. This flux choice will only excite the volume modulus of the Calabi-Yau, while the other moduli remain constant. Similarly for $F_4$ flux we can take it along $J\wedge J$ (recall that $*_6 J =\frac12 J \wedge J$). More generally, we can restrict to compactifications on Ricci flat manifolds admitting an SU(3)-structure: the resulting reduction can provide a consistent truncation, see $e.g.$~\cite{Kashani-Poor:2006ofe, Koerber:2010bx, Andriot:2018tmb}, and the SU(3)-structure forms can be used to define the isotropic flux. 
Eventually, such a flux choice creates a single exponential potential of the form
\begin{equation}
V \sim  N^2\ e^{-2(3+m)B\, \varphi +a_m\Phi}\,,   
\end{equation}
where the factor $e^{2A \varphi} = e^{-6B\varphi}$ can be understood analogously to \eqref{action4n}. As in \eqref{example1}, this is a single field exponential, where the relevant scalar is a linear combination of the two canonically normalized fields $\Phi, \varphi$. This results in the effective coupling
\begin{equation} 
\lambda^2 =  4(3+m)^2B^2 +a_m^2 = \frac{2}{3}\left(m^2 -6m +21\right)\,.    
\end{equation}
For all values $m=0,1, \dots, 6$ we have $\lambda^2>6$, meaning that we can describe it by $k=-1$ as well as $k=0$ attractors.

The attractor with $k=-1$ is given by \eqref{solPkp} for which
\begin{equation}\label{eq}
-2(3+m)B\,\varphi +a_m\Phi =-2\log t +c\,.   
\end{equation}
We can now use that, in the solution, the scalar field orthonormal to the combination appearing in the exponential is constant,
\begin{equation}\label{constant}
a_m\varphi + 2(3+m)B\, \Phi = const\,,
\end{equation}
meaning that
\begin{equation}\label{varphi}
\varphi(t) = \frac{4(3+m)B}{\lambda^2}\log t + c\,,
\end{equation}
up to a redefinition of the constant $c$. Recalling from \eqref{LKK} that the Kaluza-Klein scale is $L_{KK} \sim e^{4B\varphi} =e^{(B-A)\varphi}$, while $L_H \sim t$, we are led to
\begin{equation}\label{P}
\frac{L_{KK}}{L_H} \sim t^P\qquad\text{with}\qquad P =-\frac{m^2 -7m +18}{m^2-6m+21}<0\,.   
\end{equation}
Therefore, we always achieve scale separation as time flows for any $m$-form flux, $m=0,\ldots,6$. For the specific case $m=2$ we reproduce the example in \cite{Marconnet:2022fmx, Andriot:2023wvg}.

Let us now look at the attractors with $k=0$. They are of the kination type, namely equation \eqref{kination}. Hence we now have
\begin{equation}
-2(3+m)B\,\varphi +a_m\Phi =- |\lambda| \sqrt{\frac23}\, \log t +c\,. 
\end{equation}
Proceeding again by eliminating the orthonormal constant field, we obtain
\begin{equation}
\varphi(t) = \frac{3+m}{3\lambda}\log t + c\,.
\end{equation}
Scale separation is then calculated as before with now
\begin{equation}
P = -1 + \frac{m+3}{3 \sqrt{m(m-6) +21}}<0\ , \qquad \text{for} \qquad m=0,\dots,6.
\end{equation}
Hence, these attractors are always scale-separated.
For completeness, the values of $\lambda$ and $P$ are shown in the table below.
\begin{table}[h!]
    \centering
    \begin{tabular}{c||c | c| c | c | c | c |c}
        $m$ & 0  & 1 & 2 & 3 & 4 &5 &6  \\ \hline
        & & & & & & & \\ [-0.15in]
        $|\lambda|$ & $\sqrt{14}$ & $4\sqrt{\frac{2}{3}}$ & $\sqrt{\frac{26}{3}}$ & $2\sqrt{2}$ & $\sqrt{\frac{26}{3}}$ & $4\sqrt{\frac{2}{3}}$ & $\sqrt{14}$ \\[0.05in]
           $P({k=-1})$ & $-\frac{6}{7}$ & $-\frac{3}{4}$ &$-\frac{8}{13}$ & $-\frac12$ & $-\frac{6}{13}$ & $-\frac12$ & $-\frac47$\\[0.05in]
        $P({k=0})$ & $\frac{1}{\sqrt{21}}-1$ & $-\frac{2}{3}$ & $\frac{5}{3 \sqrt{13}}-1$ &$\frac{1}{\sqrt3}-1$ & $\frac{7}{3 \sqrt{13}}-1$ & $-\frac{1}{3}$ & $\sqrt{\frac{3}{7}}-1$
    \end{tabular}
    \label{tabPvalues}
\end{table}

These rolling solutions we have just considered are not only consistent truncations of eleven- or ten-dimensional supergravity,\footnote{The asymptotic field direction corresponds to large volume and weak coupling, $i.e.$~the supergravity regime of string theory, except for $m=6$ for which $a_m<0$ leads to strong coupling. It would then be interesting to verify whether the latter has an eleven-dimensional supergravity origin.} but also genuine four-dimensional effective descriptions. The latter becomes more and more reliable as time increases. Notice that only classical ingredients have been employed, no quantum corrections and also no orientifolds, and yet the solutions are scale-separated. The no-go theorem of \cite{Gautason:2015tig} is circumvented because these are rolling solutions and not vacua, and also because their internal curvature is vanishing.

The main conclusion is that scale separation can be achieved for rolling scenarios without orientifolds. Since asymptotic regimes of field space are characterized by runaway potentials, it does show that rolling scenarios can be completely within effective field theories. This is in sharp contrast with vacua where, in absence of difficulties to control orientifold backreaction or quantum effects, anti-de Sitter critical points are usually obtained either directly in ten/eleven dimensions or within \emph{consistent truncations}, but not within  lower-dimensional effective field theories.

\section{Entropy bounds and extra dimensions}\label{sec:entropy}

Our second main interest are entropy bounds. These can be recast in the form of inequalities involving the ultraviolet (UV) and infrared (IR) cutoff of a given effective theory; as such they are a manifestation of UV/IR mixing. Prior to discussing these bounds, we should clarify what scales we take as cutoffs.

The species scale gives an upper bound on the UV cutoff of $d$-dimensional gravitational effective theories which reads \cite{Veneziano:2001ah,Dvali:2007hz}
\begin{equation}
\label{eq:Lambdasp}
\Lambda_{sp}  \simeq \frac{M_P}{N_{sp}^{\frac{1}{d-2}}},
\end{equation}
where $N_{sp}$ is the number of (massive) species with mass below $\Lambda_{sp}$. This is the scale at which graviton loops become of similar size to the classical terms.
In a limit with $n$ decompactifying extra dimensions, and if the associated Kaluza-Klein modes are the dominant contribution to the species content, it possible to express the species scale in terms of the Kaluza-Klein scale as \cite{Castellano:2021mmx}
\begin{equation}
\label{eq:LambdaspMKK}
\Lambda_{sp} \simeq L_{KK}^{-\frac{n}{d+n-2}}M_P^{\frac{d-2}{d+n-2}}\,,
\end{equation}
which equals the $(d+n)$-dimensional Planck scale.
This formula is in principle derived in Minkowski spacetime, but \cite{Aoufia:2024awo} argue it should be valid on curved backgrounds as well, provided that $1/L_{KK} \ll M_P$. In the following, we will take $\Lambda_{sp}$ as ultraviolet cutoff as in \cite{Castellano:2021mmx}.

The question on the infrared cutoff $\Lambda_{IR}$ is more subtle. On (anti-)de Sitter vacua, it is customary to take $\Lambda_{IR} =\sqrt{|\Lambda|} \simeq 1/L_H$. This might be meaningful in de Sitter, where black holes cannot be larger than this scale, but it does not obviously appear to be well-justified in anti-de Sitter where this length scale rather sets the Hawking--Page transition. Yet, it seems the natural scale that comes out. On a rolling solution, a natural generalisation is $\Lambda_{IR} \simeq 1/L_H = H$, and we will see what it implies.  Turning the logic around, one can view entropy bounds as providing an operative definition for $\Lambda_{IR}$.

The entropy bounds we will be looking at are the Covariant Entropy Bound (CEB) \cite{Susskind:1994vu,Bousso:1999xy} and the so called CKN bound \cite{Cohen:1998zx}. 
They can be expressed as 
\begin{equation}
\label{eq:CEBandCKN}
\left(\frac{\Lambda_{UV}}{\Lambda_{IR}}\right)^\gamma \leq \left(\frac{M_P}{\Lambda_{UV}}\right)^{d-2}, \qquad \text{with} \qquad \gamma = \left\{ \begin{array}{cc}
1 & \text{CEB}\\
2 & \text{CKN}
\end{array}\right.  .
\end{equation}
The CEB encodes the holographic idea that the extensive field-theoretic entropy, $S$, of a given region cannot exceed the intensive black-hole entropy, $S_{BH}$, namely $S\leq S_{BH}$ or equivalently 
\begin{equation}
\label{eq:ceb2}
\left(\frac{\Lambda_{UV}}{\Lambda_{IR}}\right)^{d-1} \leq\left(\frac{M_P}{\Lambda_{IR}}\right)^{d-2}\, .
\end{equation}
Recently, an alternative derivation of the above bound in terms of BPS domain walls has been provided in \cite{Cribiori:2024jwq}. Consider a supersymmetric AdS vacuum with  $-L_H^{-2}M_P^{d-2} \simeq |F_d|^2$, where $F_d \simeq Q \epsilon_d$ is a $d$-dimensional electric flux whose potential $C_{d-1}$ couples to a $(d-2)$-dimensional domain wall. We assume the domain wall to be BPS, in such a way that it interpolates between supersymmetric vacua, and to be fundamental in the sense that it cannot be resolved within the $d$-dimensional effective theory. Then, its tension is $T=Q M_p^{d/2-1}$ and it provides an upper bound on the ultraviolet cutoff, $T \geq \Lambda_{UV}^{d-1}$. Combining the above information, we can write $L_H^{-2} M_P^{d-2} \simeq Q^2 \simeq T^2 M_P^{2-d}\geq \Lambda_{UV}^{2d-2}/M_P^{d-2}$, which is precisely \eqref{eq:ceb2} if $\Lambda_{IR} \simeq 1/L_H$.

The CKN bound is stronger than the CEB. The underlying idea is that the field theoretic degrees of freedom in a region of size $1/\Lambda_{IR}$ should not lie within their Schwarzschild radius. Hence, the UV cutoff should be low enough such that states with characteristic energy density $\Lambda_{UV}^d$ have Schwarzschild radius smaller than $1/\Lambda_{IR}$.  Since the mass of such  state filling the box is $\Lambda_{UV}^d/\Lambda_{IR}^{d-1}$, the associated black hole radius is found from $M_{BH}\sim L_{BH}^{d-3}M_P^{d-2}$ and so we get   
\begin{equation}
\left(\frac{\Lambda_{UV}}{\Lambda_{IR}}\right)^d \leq\left(\frac{M_P}{\Lambda_{IR}}\right)^{d-2}\, .
\end{equation}
Equivalently, in terms of entropy
\begin{equation}
S \leq S_{BH}^{\frac{d-1}{d}}.
\end{equation}

It is suggestive to derive the CKN bound from thermodynamics \cite{Ramakrishna:2021sll}. Let us consider a box of size $L_H\simeq 1/\Lambda_{IR}$ homogenously and isotropically filled with radiation with equation of state parameter $w$ such that $d=(1+w)/w$. The use of radiation instead of other fluids is motivated in order to avoid collapse inside the box. Assume the box is expanding as $L_H \simeq H^{-1}\simeq t$ with FLRW scale factor $a(t)\sim t^\frac{2}{(d-1)(1+w)}$ and energy density $\rho \sim a(t)^{(1-d)(1+w)} $. According to $d$-dimensional Stefan's law, the temperature $T$ is related to the radiation density by $\rho \sim T^d$.  
Let us consider then the free energy
\begin{equation}
    F = \rho V - TS.
\end{equation}
Here, the important point is that all quantities are given by powers of $L_H$, indeed (in $d$-dimensional Planck units)
\begin{equation}
\label{eq:LhdepCKN}
V \simeq L_H^{d-1}, \qquad \rho \simeq 1/L_H^2, \qquad T \simeq1/L_H^{2/d}.
\end{equation}
The entropy should also follow the same pattern, $S\simeq  L_H^r$, for some parameter $r$. We can now see that the value of $r$ such that the CKN bound is saturated, namely $S = S_{BH}^{(d-1)/d} \simeq  L_H^{(d-1)(d-2)/d}$ follows from extremizing the free energy with respect to $L_H$. Since $F$ is a combination of powers of $L_H$, $\partial F/\partial{L_H}  =0$ implies 
\begin{equation}
   \rho V \simeq  TS. 
\end{equation}
After substituting \eqref{eq:LhdepCKN}, we find
\begin{equation}
r = (d-1)(d-2)/d,
\end{equation}
which is the value saturating the CKN bound; for $d=4$ this is  $r=3/2$. Notice that extremization with respect to $L_H$ is basically extremization with respect to time in the argument above, hence we showed that the CKN bound is identified by extremizing the free energy with respect to time.

While the CEB is highly employed in the literature on the Swampland program, the CKN bound has also been discussed occasionally. For example, in the first derivation of the refined de Sitter conjecture \cite{Ooguri:2018wrx}, although a simpler alternative argument for it exists that does not rely on CKN \cite{Hebecker:2018vxz}, or in application to inflation \cite{Seo:2021bpb}, where it can reproduce Swampland bounds forbidding eternal inflation \cite{Rudelius:2019cfh}.

\subsection{Entropy bounds for vacua}

As a first application, we look at entropy bounds on vacua. To discuss scale separation, we are interested in setups in which species are mainly Kaluza-Klein modes. According to the ADC \cite{Lust:2019zwm}, these setups are characterized by a relation
\begin{equation}
L_{KK} \simeq L_{H}^{2\alpha}, \label{ADC}
\end{equation}
where we set $M_P=1$.
Using \eqref{eq:LambdaspMKK} and applying the entropy bounds \eqref{eq:CEBandCKN} together with \eqref{ADC}, we get \cite{Castellano:2021mmx}
\begin{align}
\alpha \geq \frac{d+n-2}{2n} \frac{\gamma}{d+\gamma -2}\qquad \text{with} \qquad \gamma = \left\{ \begin{array}{cc}
1 & \text{CEB}\\
2 & \text{CKN}
\end{array}\right.  ,
\end{align}
From \eqref{ADC}, we read that scale separation requires $\alpha < 1/2$, so that $L_{KK}<L_H$. An example is provided by the Dark Dimension \cite{Montero:2022prj} scenario with $\alpha=1/4$ and $d=4$. We now see that this scenario, which postulates $n=1$, violates both CEB and CKN. The same applies to the recently proposed double Dark Dimension \cite{Anchordoqui:2025nmb}, with $n=2$, as long as one insists on $\alpha=1/4$. In particular, for $d=4$ the special case $\alpha=1/4$ is allowed by CEB ($\gamma =1$) whenever $n\geq 4$ but never possible according to CKN ($\gamma=2$). Scale separation ($\alpha<1/2$) in $d=4$ is compatible with CEB for $n\geq 2$ and with CKN for $n\geq 3$. 

The fact that the Dark Dimension scenario violates both entropy bounds calls for a deeper understanding. 
One option is that the scenario is inconsistent. Another option is that entropy bounds apply only on vacua or, more in general, in the presence of a horizon. In this case, our analysis indicates that the Dark Dimension scenario should not be realized on a de Sitter vacuum, which does have a horizon, but instead on a rolling solution with no cosmic or apparent horizon.  Yet a third option is that the implementation of the entropy bounds we used, following \cite{Castellano:2021mmx,Calderon-Infante:2023ler}, is too naive and more refined versions are needed.\footnote{For example, by considering the UV cutoff to be the Kaluza-Klein scale, we would get $\alpha \geq \gamma /(2(d+\gamma-2))$. Then, for $d=4$ and $\alpha=1/4$ CKN would be saturated while CEB would be satisfied. In this case, when combined with $1/4 \leq \alpha \leq 1/2$ \cite{Montero:2022prj}, CKN would single out the value $\alpha =1/4$ required by the Dark Dimension scenario.} Notice however that the form of the CEB here employed has been recently re-derived in \cite{Cribiori:2024jwq} without relying on entropy arguments.

\subsection{Entropy bounds for rolling solutions}

We finally discuss the entropy bounds for the rolling solutions in the examples of section \ref{sec:vacuavsroll}. To have a weakly coupled gravitational description at the Kaluza-Klein scale, it is necessary that $\Lambda_{sp}>1/L_{KK}$. On a vacuum this is guaranteed by \eqref{eq:LambdaspMKK}, but on a rolling solution it might fail. In four-dimensional Planck units, it can be seen from \eqref{eq:LambdaspMKK} that $L_{KK}\, \Lambda_{sp}$ would be large if $L_{KK}$ grows with time, which is the case for all our solution examples.

We can now look at the entropy bounds \eqref{eq:CEBandCKN}. In units where $M_P=1$, we find
\begin{equation}
L_{KK}^{\frac{(\gamma+2)n}{n+2}} \geq t^{\gamma} .
\end{equation}
Whenever a solution is not scale separated and so $L_{KK}$ grows at least as fast as $t$, then for models from ten-dimensions ($n=6$) or eleven-dimensions ($n=7$) we satisfy both entropy bounds. 
Then, let us look at the scale-separated late time solutions where $L_{KK}/L_H\sim t^P$ with $P<0$. In this case, the entropy bounds lead to 
\begin{equation}
 (\gamma+2)(P+1)  \geq \frac{n+2}{n}\gamma\, ,   
\end{equation}
assuming $t>1$, since we cannot trust the region $t<1$. Let us consider first the solutions from eleven-dimensional supergravity, for which $n=7$. 
For the  $k=-1$ solution we have $P+1=3/7$, and thus the CEB ($\gamma=1$), is saturated whereas CKN  ($\gamma=2$) is violated. For the $k=0$ solution we have $P+1=\sqrt{3/7}$, and thus both CEB and CKN are satisfied.
For the scale separated solutions coming from ten-dimensional type II supergravity ($n=6$), we summarise the results in the table below. Notice that CKN for $k=0$ and $m=5$ is only saturated.
\begin{table}[h!]
    \centering
    \begin{tabular}{c||c | c| c | c | c | c |c}
        $m$ & 0  & 1 & 2 & 3 & 4 &5 &6  \\ \hline
        CEB ($k=-1$) & $\times$ & $\times$ & $\times$ & $\checkmark$ & $\checkmark$ & $\checkmark$  & $\times$  \\
        CKN ($k=-1$) & $\times$ & $\times$ & $\times$ & $\times$ & $\times$ & $\times$  & $\times$\\
        CEB ($k=0$) & $\times$ & $\times$ & $\checkmark$ & $\checkmark$ & $\checkmark$ & $\checkmark$  & $\checkmark$  \\
        CKN ($k=0$) & $\times$ & $\times$ & $\times$ & $\times$ & $\times$ & $\checkmark$ & $\times$
    \end{tabular}
    \label{tab:my_label2}
\end{table}

Given that these solutions do not have a cosmic horizon, it could be that these entropy bound have not been applied correctly because we used the wrong IR scale. This is compatible with the fact that these solutions have a trustworthy microscopic origin.

\section{Discussion}\label{sec:disc}

We discussed whether compactifications of string theory, within the asymptotic supergravity regime, can lead to actual lower-dimensional effective field theories. In other words, can the compactification background feature a separation between the Kaluza-Klein scale and the four-dimensional curvature scale? 
We emphasize the surprising fact that scale separation seems fairly generic for spacetimes where scalar fields are rolling, in line with the property that scalar potentials are runaway in asymptotic regimes \cite{Ooguri:2018wrx, Hebecker:2018vxz}. 
In particular, when the potential is purely generated by internal curvature we find again an obstruction to scale separation. Different is the case in which the potential is supported by fluxes, for which we provided examples of scale-separated solutions in string and M-theory. We verified this by direct computation, considering solutions to ten- and eleven-dimensional supergravity, clarifying the appropriate scales to compare, and verifying their ratio.

We then considered the issue of scale separation in light of entropy bounds; both the covariant one \cite{Bousso:1999xy} (CEB) and the stronger CKN bound \cite{Cohen:1998zx}. We note that our scale-separated rolling solutions at best saturate CKN and sometimes violate CEB. Since the solutions are trustworthy, we must conclude that either these bounds are falsified in a negative manner, or that we have used the wrong IR scale in the entropy bounds. The latter option seems in line with known violations of entropy bounds when applied too naively \cite{Bousso:2002ju} and not relying on the covariant formulation of \cite{Bousso:1999xy} using light sheets.\footnote{We thank Miguel Montero for useful discussions on this point.}  
Admittedly, it is not obvious what the IR scale is in a cosmological spacetime without cosmic horizon. 

However, when applied to (nearly) de Sitter universes there is less confusion about what IR scale and Hubble radius should be used in entropy bounds (see \cite{Fischler:1998st, Banks:2019arz, Calderon-Infante:2023ler} for a critical discussion). When we assume our own universe to be in a (meta-stable) de Sitter state we find that the Dark Dimension scenario \cite{Montero:2022prj} violates the CEB and thus also CKN, at least when implementing them following \cite{Castellano:2021mmx}. In fact, scale separation as a whole is forbidden by the CEB if there is only one extra (dark) dimension. A scenario with two extra (dark) dimensions, as proposed independently in \cite{Anchordoqui:2025nmb} and \cite{Casas:2024ttx} as a model to resolve both the Electroweak and the Cosmological hierarchy, saturates the CEB if we relax $\alpha=1/4$, but it still violates the CKN bound. We believe this can sharpen the question about the existence of top-down string theory embeddings of the Dark Dimension scenario, especially as a rolling solution rather than a vacuum \cite{Nian:2024njr}, or its version with two compact dimensions. Otherwise, it can guide us towards a more precise implementation of entropy bounds.

\paragraph{Acknowledgments.}

We thank I.~Basile, A.~Castellano, M.~Montero, I.~Ruiz, F.~Tonioni, C.~Vafa, V.~Van Hemelryck for discussions, and D.~Tsimpis for helpful exchanges on the rolling solutions. NC thanks the Harvard Swampland Initiative for hospitality while this work was under completion. 
The work of NC is supported by the Research Foundation Flanders (FWO grant 1259125N). DA thanks KU Leuven for hospitality during completion of this work, except for the served Belgian beers, by far too dangerous.

\addcontentsline{toc}{section}{References}

\small{
\bibliography{references} }

\providecommand{\href}[2]{#2}\begingroup\raggedright\begin{thebibliography}{10}

\bibitem{Vafa:2005ui}
C.~Vafa, ``{The String landscape and the swampland},''
  \href{http://www.arXiv.org/abs/hep-th/0509212}{{\tt hep-th/0509212}}.

\bibitem{Palti:2019pca}
E.~Palti, ``{The Swampland: Introduction and Review},'' {\em Fortsch. Phys.}
  {\bf 67} (2019), no.~6, 1900037,
  \href{http://www.arXiv.org/abs/1903.06239}{{\tt 1903.06239}}.

\bibitem{Agmon:2022thq}
N.~B. Agmon, A.~Bedroya, M.~J. Kang, and C.~Vafa, ``{Lectures on the string
  landscape and the Swampland},''
  \href{http://www.arXiv.org/abs/2212.06187}{{\tt 2212.06187}}.

\bibitem{Lust:2019zwm}
D.~L\"ust, E.~Palti, and C.~Vafa, ``{AdS and the Swampland},'' {\em Phys. Lett.
  B} {\bf 797} (2019) 134867, \href{http://www.arXiv.org/abs/1906.05225}{{\tt
  1906.05225}}.

\bibitem{Palti:2020tsy}
E.~Palti, ``{Fermions and the Swampland},'' {\em Phys. Lett. B} {\bf 808}
  (2020) 135617, \href{http://www.arXiv.org/abs/2005.08538}{{\tt 2005.08538}}.

\bibitem{Cribiori:2020use}
N.~Cribiori, G.~Dall'agata, and F.~Farakos, ``{Weak gravity versus de
  Sitter},'' {\em JHEP} {\bf 04} (2021) 046,
  \href{http://www.arXiv.org/abs/2011.06597}{{\tt 2011.06597}}.

\bibitem{Gonzalo:2021fma}
E.~Gonzalo, L.~E. Ib\'a\~nez, and I.~Valenzuela, ``{AdS swampland conjectures
  and light fermions},'' {\em Phys. Lett. B} {\bf 822} (2021) 136691,
  \href{http://www.arXiv.org/abs/2104.06415}{{\tt 2104.06415}}.

\bibitem{Cribiori:2021gbf}
N.~Cribiori, D.~Lust, and M.~Scalisi, ``{The gravitino and the swampland},''
  {\em JHEP} {\bf 06} (2021) 071,
  \href{http://www.arXiv.org/abs/2104.08288}{{\tt 2104.08288}}.

\bibitem{Castellano:2021yye}
A.~Castellano, A.~Font, A.~Herraez, and L.~E. Ib\'a\~nez, ``{A gravitino
  distance conjecture},'' {\em JHEP} {\bf 08} (2021) 092,
  \href{http://www.arXiv.org/abs/2104.10181}{{\tt 2104.10181}}.

\bibitem{DallAgata:2021nnr}
G.~Dall'Agata, M.~Emelin, F.~Farakos, and M.~Morittu, ``{The unbearable
  lightness of charged gravitini},'' {\em JHEP} {\bf 10} (2021) 076,
  \href{http://www.arXiv.org/abs/2108.04254}{{\tt 2108.04254}}.

\bibitem{Gonzalo:2021zsp}
E.~Gonzalo, L.~E. Ib\'a\~nez, and I.~Valenzuela, ``{Swampland constraints on
  neutrino masses},'' {\em JHEP} {\bf 02} (2022) 088,
  \href{http://www.arXiv.org/abs/2109.10961}{{\tt 2109.10961}}.

\bibitem{Cribiori:2023gcy}
N.~Cribiori and F.~Farakos, ``{Supergravity EFTs and swampland constraints},''
  {\em PoS} {\bf CORFU2022} (2023) 167,
  \href{http://www.arXiv.org/abs/2304.12806}{{\tt 2304.12806}}.

\bibitem{Casas:2024ttx}
G.~F. Casas, L.~E. Ib\'a\~nez, and F.~Marchesano, ``{Yukawa couplings at
  infinite distance and swampland towers in chiral theories},'' {\em JHEP} {\bf
  09} (2024) 170, \href{http://www.arXiv.org/abs/2403.09775}{{\tt 2403.09775}}.

\bibitem{Montero:2022prj}
M.~Montero, C.~Vafa, and I.~Valenzuela, ``{The dark dimension and the
  Swampland},'' {\em JHEP} {\bf 02} (2023) 022,
  \href{http://www.arXiv.org/abs/2205.12293}{{\tt 2205.12293}}.

\bibitem{Coudarchet:2023mfs}
T.~Coudarchet, ``{Hiding the extra dimensions: A review on scale separation in
  string theory},'' {\em Phys. Rept.} {\bf 1064} (2024) 1--28,
  \href{http://www.arXiv.org/abs/2311.12105}{{\tt 2311.12105}}.

\bibitem{Gautason:2015tig}
F.~F. Gautason, M.~Schillo, T.~Van~Riet, and M.~Williams, ``{Remarks on scale
  separation in flux vacua},'' {\em JHEP} {\bf 03} (2016) 061,
  \href{http://www.arXiv.org/abs/1512.00457}{{\tt 1512.00457}}.

\bibitem{Cribiori:2021djm}
N.~Cribiori, D.~Junghans, V.~Van~Hemelryck, T.~Van~Riet, and T.~Wrase,
  ``{Scale-separated AdS4 vacua of IIA orientifolds and M-theory},'' {\em Phys.
  Rev. D} {\bf 104} (2021), no.~12, 126014,
  \href{http://www.arXiv.org/abs/2107.00019}{{\tt 2107.00019}}.

\bibitem{VanHemelryck:2024bas}
V.~Van~Hemelryck, ``{Weak G2 manifolds and scale separation in M-theory from
  type IIA backgrounds},'' {\em Phys. Rev. D} {\bf 110} (2024), no.~10, 106013,
  \href{http://www.arXiv.org/abs/2408.16609}{{\tt 2408.16609}}.

\bibitem{Andriot:2018tmb}
D.~Andriot and D.~Tsimpis, ``{Laplacian spectrum on a nilmanifold, truncations
  and effective theories},'' {\em JHEP} {\bf 09} (2018) 096,
  \href{http://www.arXiv.org/abs/1806.05156}{{\tt 1806.05156}}.

\bibitem{Maldacena:2000mw}
J.~M. Maldacena and C.~Nunez, ``{Supergravity description of field theories on
  curved manifolds and a no go theorem},'' {\em Int. J. Mod. Phys. A} {\bf 16}
  (2001) 822--855, \href{http://www.arXiv.org/abs/hep-th/0007018}{{\tt
  hep-th/0007018}}.

\bibitem{Townsend:2003fx}
P.~K. Townsend and M.~N.~R. Wohlfarth, ``{Accelerating cosmologies from
  compactification},'' {\em Phys. Rev. Lett.} {\bf 91} (2003) 061302,
  \href{http://www.arXiv.org/abs/hep-th/0303097}{{\tt hep-th/0303097}}.

\bibitem{Emparan:2003gg}
R.~Emparan and J.~Garriga, ``{A Note on accelerating cosmologies from
  compactifications and S branes},'' {\em JHEP} {\bf 05} (2003) 028,
  \href{http://www.arXiv.org/abs/hep-th/0304124}{{\tt hep-th/0304124}}.

\bibitem{Rudelius:2022gbz}
T.~Rudelius, ``{Asymptotic scalar field cosmology in string theory},'' {\em
  JHEP} {\bf 10} (2022) 018, \href{http://www.arXiv.org/abs/2208.08989}{{\tt
  2208.08989}}.

\bibitem{Shiu:2023fhb}
G.~Shiu, F.~Tonioni, and H.~V. Tran, ``{Late-time attractors and cosmic
  acceleration},'' {\em Phys. Rev. D} {\bf 108} (2023), no.~6, 063528,
  \href{http://www.arXiv.org/abs/2306.07327}{{\tt 2306.07327}}.

\bibitem{Andriot:2023wvg}
D.~Andriot, D.~Tsimpis, and T.~Wrase, ``{Accelerated expansion of an open
  universe and string theory realizations},'' {\em Phys. Rev. D} {\bf 108}
  (2023), no.~12, 123515, \href{http://www.arXiv.org/abs/2309.03938}{{\tt
  2309.03938}}.

\bibitem{Ooguri:2018wrx}
H.~Ooguri, E.~Palti, G.~Shiu, and C.~Vafa, ``{Distance and de Sitter
  Conjectures on the Swampland},'' {\em Phys. Lett. B} {\bf 788} (2019)
  180--184, \href{http://www.arXiv.org/abs/1810.05506}{{\tt 1810.05506}}.

\bibitem{DESI:2024mwx}
{\bf DESI} Collaboration, A.~G. Adame {\em et al.}, ``{DESI 2024 VI:
  cosmological constraints from the measurements of baryon acoustic
  oscillations},'' {\em JCAP} {\bf 02} (2025) 021,
  \href{http://www.arXiv.org/abs/2404.03002}{{\tt 2404.03002}}.

\bibitem{DESI:2025zgx}
{\bf DESI} Collaboration, M.~Abdul~Karim {\em et al.}, ``{DESI DR2 Results II:
  Measurements of Baryon Acoustic Oscillations and Cosmological Constraints},''
  \href{http://www.arXiv.org/abs/2503.14738}{{\tt 2503.14738}}.

\bibitem{Castellano:2021mmx}
A.~Castellano, A.~Herr\'aez, and L.~E. Ib\'a\~nez, ``{IR/UV mixing, towers of
  species and swampland conjectures},'' {\em JHEP} {\bf 08} (2022) 217,
  \href{http://www.arXiv.org/abs/2112.10796}{{\tt 2112.10796}}.

\bibitem{Castellano:2023qhp}
A.~Castellano, A.~Herr\'aez, and L.~E. Ib\'a\~nez, ``{Towers and hierarchies in
  the Standard Model from Emergence in Quantum Gravity},'' {\em JHEP} {\bf 10}
  (2023) 172, \href{http://www.arXiv.org/abs/2302.00017}{{\tt 2302.00017}}.

\bibitem{Calderon-Infante:2023ler}
J.~Calder\'on-Infante, A.~Castellano, A.~Herr\'aez, and L.~E. Ib\'a\~nez,
  ``{Entropy bounds and the species scale distance conjecture},'' {\em JHEP}
  {\bf 01} (2024) 039, \href{http://www.arXiv.org/abs/2306.16450}{{\tt
  2306.16450}}.

\bibitem{Bousso:1999xy}
R.~Bousso, ``{A Covariant entropy conjecture},'' {\em JHEP} {\bf 07} (1999)
  004, \href{http://www.arXiv.org/abs/hep-th/9905177}{{\tt hep-th/9905177}}.

\bibitem{Cohen:1998zx}
A.~G. Cohen, D.~B. Kaplan, and A.~E. Nelson, ``{Effective field theory, black
  holes, and the cosmological constant},'' {\em Phys. Rev. Lett.} {\bf 82}
  (1999) 4971--4974, \href{http://www.arXiv.org/abs/hep-th/9803132}{{\tt
  hep-th/9803132}}.

\bibitem{Anchordoqui:2025nmb}
L.~Anchordoqui, I.~Antoniadis, and D.~Lust, ``{Two Micron-Size Dark
  Dimensions},'' \href{http://www.arXiv.org/abs/2501.11690}{{\tt 2501.11690}}.

\bibitem{Andersson:2006du}
L.~Andersson and J.~M. Heinzle, ``{Eternal acceleration from M-theory},'' {\em
  Adv. Theor. Math. Phys.} {\bf 11} (2007), no.~3, 371--398,
  \href{http://www.arXiv.org/abs/hep-th/0602102}{{\tt hep-th/0602102}}.

\bibitem{Gibbons:1984kp}
G.~W. Gibbons, ``{ASPECTS OF SUPERGRAVITY THEORIES},'' in {\em {XV GIFT Seminar
  on Supersymmetry and Supergravity}}.
\newblock 6, 1984.

\bibitem{deWit:1986mwo}
B.~de~Wit, D.~J. Smit, and N.~D. Hari~Dass, ``{Residual Supersymmetry of
  Compactified D=10 Supergravity},'' {\em Nucl. Phys. B} {\bf 283} (1987) 165.

\bibitem{VanRiet:2023pnx}
T.~Van~Riet and G.~Zoccarato, ``{Beginners lectures on flux compactifications
  and related Swampland topics},'' {\em Phys. Rept.} {\bf 1049} (2024) 1--51,
  \href{http://www.arXiv.org/abs/2305.01722}{{\tt 2305.01722}}.

\bibitem{Andriot:2022brg}
D.~Andriot, L.~Horer, and G.~Tringas, ``{Negative scalar potentials and the
  swampland: an Anti-Trans-Planckian Censorship Conjecture},'' {\em JHEP} {\bf
  04} (2023) 139, \href{http://www.arXiv.org/abs/2212.04517}{{\tt 2212.04517}}.

\bibitem{Andriot:2024jsh}
D.~Andriot, S.~Parameswaran, D.~Tsimpis, T.~Wrase, and I.~Zavala,
  ``{Exponential quintessence: curved, steep and stringy?},'' {\em JHEP} {\bf
  08} (2024) 117, \href{http://www.arXiv.org/abs/2405.09323}{{\tt 2405.09323}}.

\bibitem{Lucchin:1984yf}
F.~Lucchin and S.~Matarrese, ``{Power Law Inflation},'' {\em Phys. Rev. D} {\bf
  32} (1985) 1316.

\bibitem{vandenHoogen:1999qq}
R.~J. van~den Hoogen, A.~A. Coley, and D.~Wands, ``{Scaling solutions in
  Robertson-Walker space-times},'' {\em Class. Quant. Grav.} {\bf 16} (1999)
  1843--1851, \href{http://www.arXiv.org/abs/gr-qc/9901014}{{\tt
  gr-qc/9901014}}.

\bibitem{Marconnet:2022fmx}
P.~Marconnet and D.~Tsimpis, ``{Universal accelerating cosmologies from 10d
  supergravity},'' {\em JHEP} {\bf 01} (2023) 033,
  \href{http://www.arXiv.org/abs/2210.10813}{{\tt 2210.10813}}.

\bibitem{Andriot:2020lea}
D.~Andriot, N.~Cribiori, and D.~Erkinger, ``{The web of swampland conjectures
  and the TCC bound},'' {\em JHEP} {\bf 07} (2020) 162,
  \href{http://www.arXiv.org/abs/2004.00030}{{\tt 2004.00030}}.

\bibitem{Casas:2024oak}
G.~F. Casas and I.~Ruiz, ``{Cosmology of light towers and swampland
  constraints},'' {\em JHEP} {\bf 12} (2024) 193,
  \href{http://www.arXiv.org/abs/2409.08317}{{\tt 2409.08317}}.

\bibitem{Kashani-Poor:2006ofe}
A.-K. Kashani-Poor and R.~Minasian, ``{Towards reduction of type II theories on
  SU(3) structure manifolds},'' {\em JHEP} {\bf 03} (2007) 109,
  \href{http://www.arXiv.org/abs/hep-th/0611106}{{\tt hep-th/0611106}}.

\bibitem{Koerber:2010bx}
P.~Koerber, ``{Lectures on Generalized Complex Geometry for Physicists},'' {\em
  Fortsch. Phys.} {\bf 59} (2011) 169--242,
  \href{http://www.arXiv.org/abs/1006.1536}{{\tt 1006.1536}}.

\bibitem{Veneziano:2001ah}
G.~Veneziano, ``{Large N bounds on, and compositeness limit of, gauge and
  gravitational interactions},'' {\em JHEP} {\bf 06} (2002) 051,
  \href{http://www.arXiv.org/abs/hep-th/0110129}{{\tt hep-th/0110129}}.

\bibitem{Dvali:2007hz}
G.~Dvali, ``{Black Holes and Large N Species Solution to the Hierarchy
  Problem},'' {\em Fortsch. Phys.} {\bf 58} (2010) 528--536,
  \href{http://www.arXiv.org/abs/0706.2050}{{\tt 0706.2050}}.

\bibitem{Aoufia:2024awo}
C.~Aoufia, I.~Basile, and G.~Leone, ``{Species scale, worldsheet CFTs and
  emergent geometry},'' \href{http://www.arXiv.org/abs/2405.03683}{{\tt
  2405.03683}}.

\bibitem{Susskind:1994vu}
L.~Susskind, ``{The World as a hologram},'' {\em J. Math. Phys.} {\bf 36}
  (1995) 6377--6396, \href{http://www.arXiv.org/abs/hep-th/9409089}{{\tt
  hep-th/9409089}}.

\bibitem{Cribiori:2024jwq}
N.~Cribiori, F.~Farakos, and N.~Liatsos, ``{On scale-separated supersymmetric
  AdS$_2$ flux vacua},'' \href{http://www.arXiv.org/abs/2411.04932}{{\tt
  2411.04932}}.

\bibitem{Ramakrishna:2021sll}
S.~Ramakrishna, ``{A thermodynamic origin for the Cohen-Kaplan-Nelson bound},''
  {\em EPL} {\bf 136} (2021), no.~3, 31001,
  \href{http://www.arXiv.org/abs/2111.07807}{{\tt 2111.07807}}.

\bibitem{Hebecker:2018vxz}
A.~Hebecker and T.~Wrase, ``{The Asymptotic dS Swampland Conjecture - a
  Simplified Derivation and a Potential Loophole},'' {\em Fortsch. Phys.} {\bf
  67} (2019), no.~1-2, 1800097, \href{http://www.arXiv.org/abs/1810.08182}{{\tt
  1810.08182}}.

\bibitem{Seo:2021bpb}
M.-S. Seo, ``{Implication of the swampland distance conjecture on the
  Cohen\textendash{}Kaplan\textendash{}Nelson bound in de Sitter space},'' {\em
  Eur. Phys. J. C} {\bf 82} (2022), no.~4, 338,
  \href{http://www.arXiv.org/abs/2106.00138}{{\tt 2106.00138}}.

\bibitem{Rudelius:2019cfh}
T.~Rudelius, ``{Conditions for (No) Eternal Inflation},'' {\em JCAP} {\bf 08}
  (2019) 009, \href{http://www.arXiv.org/abs/1905.05198}{{\tt 1905.05198}}.

\bibitem{Bousso:2002ju}
R.~Bousso, ``{The Holographic principle},'' {\em Rev. Mod. Phys.} {\bf 74}
  (2002) 825--874, \href{http://www.arXiv.org/abs/hep-th/0203101}{{\tt
  hep-th/0203101}}.

\bibitem{Fischler:1998st}
W.~Fischler and L.~Susskind, ``{Holography and cosmology},''
  \href{http://www.arXiv.org/abs/hep-th/9806039}{{\tt hep-th/9806039}}.

\bibitem{Banks:2019arz}
T.~Banks and P.~Draper, ``{Remarks on the Cohen-Kaplan-Nelson bound},'' {\em
  Phys. Rev. D} {\bf 101} (2020), no.~12, 126010,
  \href{http://www.arXiv.org/abs/1911.05778}{{\tt 1911.05778}}.

\bibitem{Nian:2024njr}
G.-E. Nian and S.~Vandoren, ``{Towards a String Realization of the Dark
  Dimension via T-folds},'' \href{http://www.arXiv.org/abs/2411.19216}{{\tt
  2411.19216}}.

\end{thebibliography}\endgroup

\bibliographystyle{utphys}

\end{document}